\documentclass[doublecol]{epl2} 
\usepackage{amsmath,amssymb}
\usepackage{graphics,epsfig}
\usepackage{graphicx}
\usepackage{color}

\def\build#1_#2^#3{\mathrel{\mathop{\kern 0pt#1}\limits_{#2}^{#3}}}

 \newcommand{\vct}[1]{{\mbox {\boldmath $#1$}}}

\title{A Stochastic Representation of the Local Structure of Turbulence}
\shorttitle{A Stochastic Representation of the Local Structure of Turbulence} 

\author{Laurent Chevillard\inst{1}, Raoul Robert\inst{2} and Vincent Vargas\inst{3}}
\shortauthor{L. Chevillard \etal}

\institute{                    
  \inst{1} Laboratoire de Physique de l'ENS Lyon, CNRS, Universit\'e de Lyon, 46 all\'ee d'Italie, 69007 Lyon, France\\
  \inst{2} Institut Fourier, CNRS, Universit\'e Grenoble 1, 100 rue des Math\'ematiques, BP 74, 38402 Saint-Martin d'H\`eres cedex, France\\
\inst{3} Ceremade, CNRS, Universit\'e Paris-Dauphine, F-75016 Paris, France.
}
\pacs{47.27.Gs}{Isotropic turbulence}
\pacs{02.50.Ey}{Stochastic processes}
\pacs{47.53.+n}{Fractals in fluid dynamics}

\abstract{
Based on the mechanics of the Euler equation at short time, we show that a Recent Fluid Deformation (RFD) closure for the vorticity field, neglecting the early stage of advection of fluid particles, allows to build a 3D incompressible velocity field that shares many properties with empirical turbulence, such as the teardrop shape of the R-Q plane. Unfortunately, non gaussianity is weak (i.e. no intermittency) and vorticity gets preferentially aligned with the wrong eigenvector of the deformation. {We then show that slightly modifying the former vectorial field in order to impose the long range correlated nature of turbulence allows to reproduce the main properties of stationary flows}. Doing so, we end up with a realistic incompressible, skewed and intermittent velocity field that reproduces the main characteristics of 3D turbulence in the inertial range, including correct vorticity alignment properties.
}

\begin{document}

\maketitle

\section{Introduction}

Fully developed turbulent flows are omnipresent in Nature ({e.g.} meteorology) and engineering ({e.g.} combustion). Despite an apparent complexity, it turns out that these flows exhibit 
universal statistical properties such as the Kolmogorov $k^{-5/3}$-law observed on the power spectrum \cite{K41} and the intermittent nature of longitudinal and transverse velocity fluctuations \cite{Fri95}. Modern {developments} of experimental and numerical facilities \cite{Tsino01,Wal09} have furthermore underlined the peculiar and universal geometry of turbulence.

From a mathematical viewpoint, our knowledge of Navier-Stokes and Euler equations is still poor and this makes the situation far from clear if we look for a relevant model of turbulent flows. {Indeed, one can prove the existence of dissipative solutions of Euler equations and the associated inertial dissipation process has been studied \cite{DucRob00}. As a consequence the introduction of viscosity is not the only way we have to build realistic model of turbulent flows.} Some progresses have been achieved recently in the understanding of the universality of the small scales of turbulence {by} studying the Lagrangian dynamics of the velocity gradient tensor $A_{ij} = \partial_ju_i$, where $\textbf{u}$ denotes the velocity (see \cite{CheA2} and references therein). While this approach takes completely into account the local interactions governing the dynamics of $\textbf{A}$, it requires closures for both the pressure Hessian and viscous term. As far as we know, provided closures miss at least partially the nonlocal nature of pressure \cite{CheA2}. An alternative approach would be devoted to consider the spatial distribution of the vectorial velocity field. Very few theoretical works have focused on this difficult, although important, aspect. Nonetheless, it has been shown that taking into account numerically the short time advection of fluid particles for all scales of motion allows to build a realistic velocity field that shares many properties {with} empirical turbulence \cite{RosMen}. Unfortunately, incompressibility has to be imposed at every scale, leading to a complicated construction method of this field, making it not fully explicit. {Another route would be devoted to directly proposing a wide class of intermittent vectorial fields \cite{RobVar08}. But the main limitation of such models is that the distribution law of the velocity increments is symmetric, so that (following the 4/5 Kolmogorov law) such fields do not exhibit energy dissipation. As it is discussed in \cite{RobVar08},  it appears that it is a very intricate issue to construct an incompressible random field with dissymmetric increments and non zero dissipation. This issue is the subject of the proposed article and while in the symmetric case explicit analytical calculations can be performed, they are out of reach in the present case and consequently we make use of numerical simulations to study the model.}
 
In this letter, based on former works \cite{CheA2,RobVar08}, we propose a stochastic method to build  an incompressible, skewed and intermittent velocity field. This method is motivated by the early stage mechanics of the Euler equation during which vorticity is stretched by the local deformation, whereas early advection by the large scale velocity is neglected. We will see that such a Recent Fluid Deformation (RFD)  closure \cite{CheA2} leads to an incompressible differentiable velocity field which reproduces well known facts of empirical turbulence, namely the teardrop shape of the RQ plane and a skewed probability density function (PDF) for the longitudinal gradients. Unfortunately, it is shown numerically that this field is not skewed in the inertial range, leading to vanishing mean energy transfer through scales, and furthermore, alignment properties of vorticity deviate from empirical findings. {We then show that a slight modification of this field, inspired by the multifractal phenomenology, is able to reproduce the main properties of stationary turbulence.}

\section{Recent fluid deformation closure}

Let us now introduce a flavor of Euler dynamics in the picture. The Euler equation writes:
\begin{equation}\label{eq:Euler}
 \left\{
\begin{array}{ll}
\frac{\partial \textbf{u}}{\partial t} + (\textbf{u}.\vct{\nabla})\textbf{u} = -\vct{\nabla}p\\
\vct{\nabla}.\textbf{u}=0
\end{array}
\right.\end{equation}
It is classical to introduce the vorticity field $\vct{\omega} (t,\textbf{x}) = \vct{\nabla}\wedge \textbf{u} (t,\textbf{x})$, and take the curl of (\ref{eq:Euler}) to eliminate the pressure and get the Beltrami equation:
\begin{equation}\label{eq:Beltrami}
\frac{\partial \vct{\omega}}{\partial t} = (\vct{\nabla} \textbf{u})\vct{\omega}-(\textbf{u}.\vct{\nabla})\vct{\omega}\mbox{ ,}
\end{equation}
which together with the system $\vct{\nabla}\wedge  \textbf{u} = \vct{\omega}$ and $\vct{\nabla}.\textbf{u}=0$
gives a closed equation in $\vct{\omega} (t,\textbf{x}) $. If vorticity vanishes at infinity, the solution of this system is given by the classical Biot-Savart formula:
\begin{equation}\label{eq:BiotSavard}
\textbf{u}(t,\textbf{x})= -\frac{1}{4\pi} \int \frac{\textbf{x}-\textbf{y}}{|\textbf{x}-\textbf{y}|^3}\wedge \vct{\omega}(t,\textbf{y}) d\textbf{y}\mbox{ .}
\end{equation}
In what follows, we shall suppose that we have a smooth solution of the system (\ref{eq:Beltrami}), (\ref{eq:BiotSavard}) with initial data $\vct{\omega}_0$. Then, it will be convenient to introduce the associated Lagrangian flow $\textbf{X}(t,\textbf{x})$ defined by the ordinary differential equation  $\frac{d\textbf{X}(t,\textbf{x})}{dt} = \textbf{u}(t,\textbf{X}(t,\textbf{x}))$ and $\textbf{X}(0,\textbf{x})=\textbf{x}$.
Using $\textbf{X}(t,\textbf{x})$, it is easy to see that (\ref{eq:Beltrami}) then writes $ \frac{d\vct{\omega}(t,\textbf{X}(t,\textbf{x}))}{dt} =  (\vct{\nabla} \textbf{u})\vct{\omega}(t,\textbf{X}(t,\textbf{x}))$
or equivalently: 
\begin{equation}\label{eq:AdvecWS}
\frac{d\vct{\omega}(t,\textbf{X}(t,\textbf{x}))}{dt} = \textbf{S}\vct{\omega}(t,\textbf{X}(t,\textbf{x}))
\end{equation}
where $\textbf{S}$ is the deformation rate tensor defined by the splitting of the tensor $\vct{\nabla} \textbf{u}$ into antisymmetric and symmetric parts $\vct{\nabla} \textbf{u} = \frac{1}{2}\vct{\omega}\wedge .+\textbf{S}$. Let us now focus on the short time evolution of the system (\ref{eq:AdvecWS}). Since we suppose that the solution is regular, we can linearize (\ref{eq:AdvecWS}) in the neighborhood of zero, replacing thus $\textbf{S}$ by $\textbf{S}_0$ (the strain associated to the initial vorticity $\vct{\omega}_0$), which gives:
\begin{equation}\label{eq:RFDOmega}
\vct{\omega}(t,\textbf{x})\approx e^{t\textbf{S}_0}\vct{\omega}_0(\textbf{x}-t\textbf{u}_0 (\textbf{x}))\mbox{ ,}
\end{equation}
using the fact that $\textbf{X}(t,\textbf{x}) \approx \textbf{x}+t\textbf{u}_0 (\textbf{x})$. In a first step, we will neglect the advection of the vorticity by the velocity field and only consider the stretching of the vorticity by the initial strain tensor $\textbf{S}_0$, which gives at time t:
\begin{equation}\label{eq:RFDtoU}
\textbf{u}(t,\textbf{x})= -\frac{1}{4\pi} \int \frac{\textbf{x}-\textbf{y}}{|\textbf{x}-\textbf{y}|^3}\wedge e^{t\textbf{S}_0(\textbf{y})}\vct{\omega}_0(\textbf{y})d\textbf{y}\mbox{ .}
\end{equation}
Starting with the Biot-Savart formula, classical calculations  \cite{ConsMajd,ConsMajd_2} give:
\begin{align}\label{eq:PVS}
\textbf{S}_0 (\textbf{y})&= \frac{3}{8\pi}\mbox{P.V.}\int\left[\frac{(\textbf{y}-\vct{\sigma})\otimes [(\textbf{y}-\vct{\sigma})\wedge \vct{\omega}_0(\vct{\sigma})]}{|\textbf{y}-\vct{\sigma}|^5}\right. \notag \\
& + \left.\frac{ [(\textbf{y}-\vct{\sigma})\wedge \vct{\omega}_0(\vct{\sigma})]\otimes(\textbf{y}-\vct{\sigma})}{|\textbf{y}-\vct{\sigma}|^5}\right]d\vct{\sigma}\mbox{ ,}
\end{align}
where the integral is understood as a Cauchy Principal Value (P.V.) and $\otimes$ the tensor product, i.e. $\textbf{x}\otimes\textbf{y} = x_iy_j$. Now, it is tempting to introduce in formula (\ref{eq:RFDtoU}) a random field $\vct{\omega}_0$ which is divergence-free, homogeneous, isotropic, Gaussian and with K41 scaling, that is formally \cite{RobVar08}:
\begin{equation}
\vct{\omega}_0(\textbf{x}) =  \int \frac{\textbf{x}-\textbf{y}}{|\textbf{x}-\textbf{y}|^{\frac{3}{2}+\frac{2}{3}+1}}\wedge d\textbf{W}(\textbf{y})\mbox{ ,}\notag
\end{equation}
where $d\textbf{W}(\textbf{y}) = (dW_1(\textbf{y}),dW_2(\textbf{y}),dW_3(\textbf{y}))$ is the standard vector white noise on $\mathbb R^3$. A more straightforward way to do this is to take for $\vct{\omega}_0(\textbf{y})$ the white noise $d\textbf{W}(\textbf{y})$ and only change to appropriate values the exponents of the denominators in the kernels giving $\textbf{u}(\textbf{x})$ ($|\textbf{x}-\textbf{y}|^{-3}$ is replaced by $|\textbf{x}-\textbf{y}|^{-(\frac{3}{2}+\frac{2}{3})}$) and $\textbf{S}_0(\textbf{x})$ ($|\textbf{x}-\textbf{y}|^{-5}$  by $|\textbf{x}-\textbf{y}|^{-\beta}$, with $\beta = 2+\frac{3}{2}+\frac{2}{3}$). Notice that now, the integral in the modified (\ref{eq:PVS}) is no more a principal value. These considerations lead finally to define the random field:
\begin{equation}\label{eq:K41Field}
\textbf{u}(t,\textbf{x})= -\frac{1}{4\pi} \int \frac{\textbf{x}-\textbf{y}}{|\textbf{x}-\textbf{y}|_\epsilon^{\frac{3}{2}+\frac{2}{3}}}\varphi_L (\textbf{x}-\textbf{y})\wedge e^{t\textbf{S}_0(\textbf{y})}d\textbf{W}(\textbf{y})\mbox{ ,}
\end{equation}
with
\begin{align}
\textbf{S}_0 (\textbf{y})&= \frac{3}{8\pi}\int\left[\frac{(\textbf{y}-\vct{\sigma})\otimes [(\textbf{y}-\vct{\sigma})\wedge d\textbf{W}(\vct{\sigma})]}{|\textbf{y}-\vct{\sigma}|_\epsilon^\beta}\right.\notag \\
& \left.+ \frac{ [(\textbf{y}-\vct{\sigma})\wedge d\textbf{W}(\vct{\sigma})]\otimes(\textbf{y}-\vct{\sigma})}{|\textbf{y}-\vct{\sigma}|_\epsilon^\beta}\right]\varphi_L (\textbf{y}-\vct{\sigma})\mbox{ ,}\notag
\end{align}
where, in order to get mathematically well defined integrals, we have introduced both a large scale cutoff $\varphi_L (\textbf{x}-\textbf{y})$ in the definition of $\textbf{u}$ and  $\textbf{S}_0$, and a small scale regularization $\epsilon$: $|\textbf{x}|_\epsilon = \theta_\epsilon * |\textbf{x}|$ ($*$ stands for the convolution product) , where $\theta$ is a radially symmetrical function, such that $\int_{\mathbb R ^3} \theta (\textbf{x})d\textbf{x}=1$ and $ \theta_\epsilon (\textbf{x})= \frac{1}{\epsilon^3}\theta (\textbf{x}/\epsilon)$\cite{RobVar08}.

\begin{figure}[t]
\epsfig{file=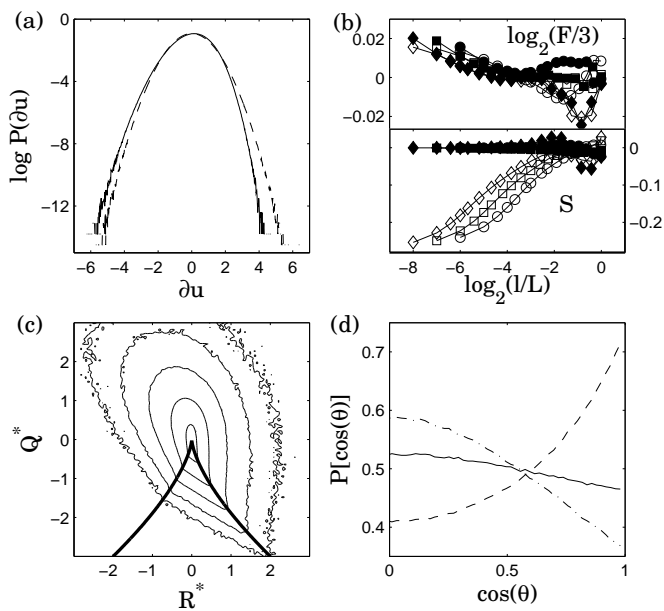,width=8.8cm}
\caption{Numerical simulations of the process given in (\ref{eq:K41Field}). (a) PDF of longitudinal (solid line) and transverse (dashed line) velocity gradients for the $N=512$ case. (b) Skewness (S) and flatness (F) of longitudinal (open symbols) and transverse (filled symbols) for the three resolutions: $N=128$ ($\circ$), $N=256$ ($\square$) and $N=512$ ($\diamond$). (c) Contour plots of the logarithm of the joint probability of the two invariants of $\textbf{A}$ ($N=512$ case) non-dimensionalized by the average strain $Q^*=Q/\langle S_{ij}S_{ij}\rangle$ and $R^*=R/\langle S_{ij}S_{ij}\rangle^{3/2}$. The thick line corresponds to the zero discriminant (Vieillefosse) line. Contour lines correspond to probabilities $10^{-4}, 10^{-3},10^{-2},10^{-1}, 1$. (d) PDF of the cosine of the angle $\theta$ between vorticity and the eigenvectors of the strain (see text) associated to three eigenvalues $\lambda_1$ (dashed-dot), $\lambda_2$ (solid) and $\lambda_3$ (dashed).}
\label{fig:K41}
\end{figure}

\section{Numerical study} 

We would like now to study the statistical properties of the velocity field defined by Eq. (\ref{eq:K41Field}). Analytical formulas are difficult to obtain, thus we will focus on numerical simulations. To do so, one has to choose the short time scale $t=\tau$. It is easy to check that the variance of the matrix $\textbf{S}_0$ (defined as $\langle \mbox{tr } \textbf{S}_0^2\rangle$) goes to infinity as the small scale parameter $\epsilon$ goes to zero. So we take for $\tau$ the local normalizing value $\tau =  (\mbox{tr } \textbf{S}_0^2)^{-1/2}$, in the spirit of the RFD closures provided in Ref. \cite{CheA2}. 

The simulation is performed in a 1-periodic box with $N^3$ collocation points. The infinitesimal volume is given by $dV=dx^3$, with $dx=1/N$. We choose as a regularizing function and large-scale cut-off the isotropic normalized Gaussian function $\varphi_{L} (\textbf{x}) = \theta_{L}(\textbf{x}) = \left(\frac{6}{\pi L^2}\right)^{3/2}\exp (-6|\textbf{x}|^2/L^2)$. This allows to compute analytically the regularized norm of a vector $\textbf{x}$, useful for numerical purposes: $|\textbf{x}|_\epsilon = \frac{\epsilon}{\sqrt{6\pi}}e^{-6\frac{|\textbf{x}|^2}{\epsilon^2}} + \left(|\textbf{x}| + \frac{\epsilon^2}{12|\textbf{x}|}\right)\mbox{erf}\left( \frac{\sqrt{6}|\textbf{x}|}{\epsilon}\right)\mbox{ ,}$
with erf the error function. 
The small-scale cut-off is chosen as $\epsilon=2dx$ and the large one as $L=1/2$.  The kernels of the form $\textbf{x}/|\textbf{x}|_\epsilon^a$ entering in Eq. (\ref{eq:K41Field}), with $a=\frac{3}{2}+\frac{2}{3}$ or $a=\beta$, are estimated in the physical space in a periodic fashion. White noise components $dW_i$, of zero mean and of variance $dV$, are generated in the physical space using a  standard random Gaussian generator. Convolution products are then performed in the Fourier space. The matrix exponential is evaluated at each point of space using a Pad\'e approximant with scaling and squaring \cite{MolLoa03}. We choose $N=128, 256, 512$. Results are displayed in Fig. \ref{fig:K41}. 

In Fig.  \ref{fig:K41}(a), we represent the longitudinal and transverse velocity gradient PDFs for the $N=512$ case. We see indeed that the longitudinal PDF is skewed, but not the transverse one (for symmetry reasons). To further characterize the structure in scale of this velocity field, we represent in Fig.  \ref{fig:K41}(b), the dependence on the scale $\ell$ of the Skewness $S=\langle (\delta_\ell u)^3\rangle/\langle (\delta_\ell u)^2\rangle^{3/2}$ and Flatness $F=\langle (\delta_\ell u)^4\rangle/\langle (\delta_\ell u)^2\rangle^{2}$ of the velocity increments $\delta_\ell u = u(x+\ell) - u(x)$, in both the longitudinal (open symbols) and transverse (filled symbols) cases. We see that S vanishes and F is consistent with a Gaussian process (i.e. $F=3$) in the inertial range. This means that the weak non-Gaussianity observed on the velocity gradients does not survive in the inertial range for the velocity increments. To further characterize the local structure of this field, we represent the joint probability of two important invariants of the velocity gradient tensor, namely $Q=-\frac{1}{2} \mbox{tr}(\textbf{A}^2)$ and $R=-\frac{1}{3}\mbox{tr}(\textbf{A}^3)$. This so-called RQ-plane has been extensively studied experimentally and numerically (see \cite{Tsino01,Wal09} for comparisons). As in empirical data, the RQ-plane is elongated along the right-tail of the Vieillefosse line, showing predominance of both enstrophy-enstrophy production (upper-left quadrant) and dissipation-dissipation production (lower-right) regions. Finally, an important nontrivial property of 3D turbulence is the preferential alignments of vorticity with the intermediate eigenvector of the deformation  \cite{Tsino01,Wal09}. We represent in  \ref{fig:K41}(d), the probability density of the cosine  of the angle between vorticity and the eigenvectors $\textbf{e}_{\lambda_i}$ of the deformation $\theta = (\vct{\omega},\textbf{e}_{\lambda_i})$ with $\lambda_1<\lambda_2<\lambda_3$. We first see that the vorticity is preferentially orthogonal to $\textbf{e}_{\lambda_1}$, as in empirical data \cite{Tsino01}. It has been observed that the vorticity gets preferentially aligned with $\textbf{e}_{\lambda_2}$, as modeled in \cite{CheA2}. We see in Fig. \ref{fig:K41}(d) that the opposite is observed in our synthetic field, namely, vorticity gets preferentially aligned with  $\textbf{e}_{\lambda_3}$. {In the following, we will see that including multifractality will allow us to predict, among other features, correct alignments.}

\section{Including multifractality}

In the first part, we have deformed at short times a K41 incompressible Gaussian field by the \textit{deformation} part of the Euler flow. As we have seen, such a velocity field (Eq. (\ref{eq:K41Field})) is not intermittent and moreover, velocity fluctuations are not skewed in the inertial range, i.e. there is no mean energy transfer across scales. A first idea would be to iterate the construction of this field several times and look for a fixed point if it exists. Preliminary simulations {indicate} that iterating this {construction} makes intermittency grow, although it is not clear if the obtained field is scale invariant and still, vorticity alignments are not correct (data not shown). An interesting development would be to apply this construction at each scale of the flow, in the spirit of Ref. \cite{RosMen}.  This remains to be explored. Anyway, at this stage, we are missing a basic property of turbulence, namely the existence of long range correlations. Indeed, the Russian school showed that the dissipation field is correlated over the large integral length scale of the flow, as it has been found experimentally (see \cite{MonYag75} and references therein). Similar observations have been made on the acceleration in Lagrangian turbulence \cite{MorDel02,YeuMordant,YeuMordant_2}: acceleration is correlated over the Kolmogorov time scale, whereas its magnitude is correlated over the integral time scale. In particular, it has been proposed in \cite{MorDel02} a multifractal random walk able to reproduce this very peculiar property. We propose in this part a generalization of this 1D intermittent model of turbulence consistent with the vectorial field structure of Eulerian turbulence.

Following previous works \cite{Multi,Multi_2,Multi_3,Multi_4,Multi_5,Multi_6}, it has been proposed in \cite{RobVar08} general ideas leading to intermittent vectorial fields that reproduce the correlation structure of the dissipation field. The main underlying idea behind building up an intermittent (scalar or vectorial) field is to take the exponential of a noise correlated logarithmically over a large scale $L$. For turbulence, a way to achieve this is to consider incompressible homogeneous, isotropic and intermittent velocity fields of the form
\begin{equation}\label{eq:notskewu}
 \textbf{u}_\epsilon(\textbf{x}) =-\frac{1}{4\pi} \int \varphi_L (\textbf{x}-\textbf{y}) \frac{\textbf{x}-\textbf{y}}{|\textbf{x}-\textbf{y}|_\epsilon^{\frac{3}{2}+\frac{2}{3}}}\wedge e^{ X_\epsilon(\textbf{y})-C_\epsilon}d\textbf{W}(\textbf{y})
\end{equation}
where the scalar field $X_\epsilon (\textbf{y})$ is defined by the following scalar product
\begin{equation}\label{eq:Xscal}
X_\epsilon (\textbf{y}) = \lambda\int_{|\textbf{y}-\vct{\sigma}|\le L} \frac{\textbf{y}-\vct{\sigma}}{|\textbf{y}-\vct{\sigma}|_\epsilon^{\frac{3}{2}+1}}.d\textbf{W'}(\textbf{\vct{\sigma}})\mbox{ .}
\end{equation}
Here, the vectorial white noises $d\textbf{W}$ and $d\textbf{W'}$ are independent. The constant $C_\epsilon$ is chosen such that the velocity field converges to a non trivial field $\textbf{u}$ as $\epsilon$ goes to zero \cite{RobVar08}. It can be shown rigorously that this vectorial field is intermittent with structure function exponents, i.e. $M_q^L(|\textbf{r}|)=\langle |(\textbf{u}(\textbf{x}+\textbf{r}) - \textbf{u}(\textbf{x})).\frac{\textbf{r}}{|\textbf{r}|}|^q\rangle \sim |r|^{\zeta_q^L}$ in the longitudinal case and $M_q^T(|\textbf{r}|)=\langle |(\textbf{u}(\textbf{x}+\textbf{r}) - \textbf{u}(\textbf{x}))\wedge\frac{\textbf{r}}{|\textbf{r}|}|^q\rangle \sim |r|^{\zeta_q^T}$ in the transverse one, behaving as a quadratic function of the order $q$, namely $\zeta_q^L=\zeta_q^T = q/3-2\pi\lambda^2q(q-2)$.  The free parameter $\lambda$ entering in the definition of the scalar field $X_\epsilon$ (Eq. (\ref{eq:Xscal})) is called the intermittency coefficient. It can be easily shown in the limit $\epsilon \ll |\textbf{y}|\rightarrow 0$ that $\langle X_\epsilon(\textbf{y})X_\epsilon(\vct{0})\rangle \sim 4\pi\lambda^2 \ln(L/|\textbf{y}|)$. We are indeed taking in the model given by the formula Eqs. (\ref{eq:notskewu}) and (\ref{eq:Xscal}) the exponential a Gaussian noise correlated over the integral length scale $L$ in a logarithmic fashion.   Unfortunately, at this stage, for symmetry reasons, the velocity increments do not exhibit asymmetry and thus, their skewness vanishes for any scale. However, it is proposed in \cite{RobVar08} another compressible version of this velocity field that is shown to provide skewness as soon the very same noise $d\textbf{W}$ entering in Eqs. (\ref{eq:notskewu}) and (\ref{eq:Xscal}) was used in the construction. It still remains to build up an incompressible and skewed intermittent velocity field that reproduces furthermore non trivial geometrical properties such as the asymmetry of the RQ plane and the alignments of vorticity with the eigenframe of the dissipation. To do that, one has to consider, as it is suggested by the early time advection of fluid particles (Eq. (\ref{eq:K41Field})), the exponentiation of a tensor field instead of a simple scalar field.

\begin{figure}[t]
\begin{center}
\epsfig{file=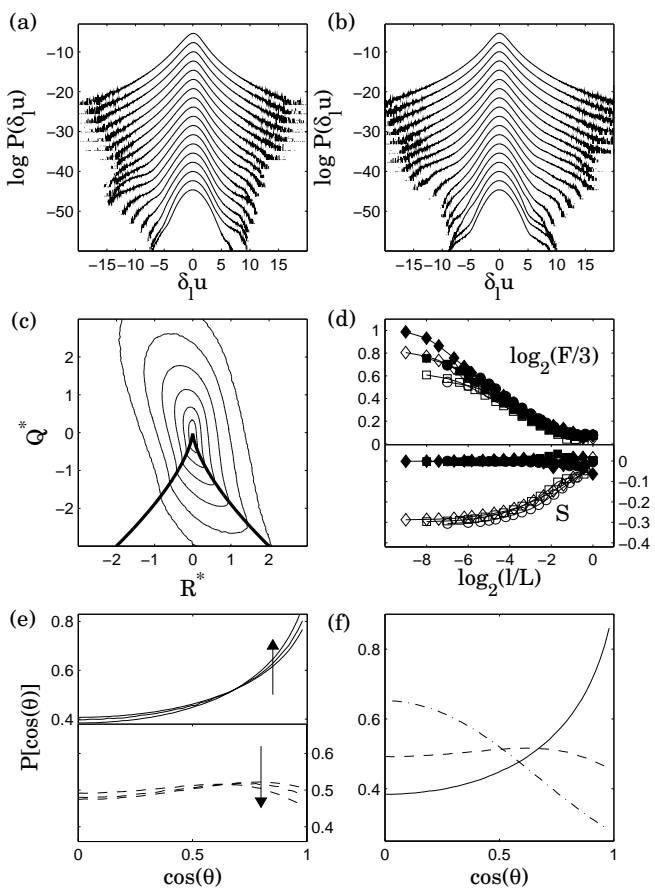,width=8.8cm}
\end{center}
\caption{\label{fig:KO62} Numerical results of the process given in (\ref{eq:KO62Field}). PDFs of longitudinal (a) and transverse (b) velocity increments $\delta_\ell u$ ($N=1024$), scales $\ell$ are logarithmically spaced between $dx$ and $L$. (c) RQ-plane, as in Fig. \ref{fig:K41}(c), for $N=1024$, contour lines correspond to probabilities $10^{-2.5}, 10^{-2},   10^{-1.5}, 10^{-1}, 10^{-.5}, 1$. Scale dependence of the Skewness (S) and Flatness (F) of the velocity increments, as in Fig. \ref{fig:K41}(b): $N=256$ ($\circ$), $N=512$ ($\square$) and $N=1024$ ($\diamond$). (e) PDF of the cosine of the angle $\theta$ between vorticity and the eigenvectors of the strain $\textbf{e}_{\lambda_2}$ (top) and $\textbf{e}_{\lambda_3}$ (bottom) for the three resolutions. The arrows indicate increasing $N$. (f) PDF of the cosine of the angle $\theta$ between vorticity and the eigenvectors of the strain as in Fig. \ref{fig:K41}(d) for $N=1024$.}
\end{figure}

Motivated by the RFD approach of the first part of this article and the explicit  intermittent velocity field shown in Eqs. (\ref{eq:notskewu}) and (\ref{eq:Xscal}), we choose to modify directly the field given in (\ref{eq:K41Field}) in order to get a multifractal velocity field. To do so, one needs to introduce this unknown intermittency parameter $\lambda$ and to change the exponent $\beta$ entering in the associated strain $\textbf{S}_0$ of (\ref{eq:K41Field}) to $\beta=3/2 +2 = 7/2$ in order to impose logarithmic long range correlations over the integral length scale $L$. Accordingly, we consider the incompressible field:
\begin{equation} 
\label{eq:KO62Field}
\widetilde{\textbf{u}}_\epsilon(\textbf{x}) =-\frac{1}{4\pi} \int \varphi_L (\textbf{x}-\textbf{y}) \frac{\textbf{x}-\textbf{y}}{|\textbf{x}-\textbf{y}|_\epsilon^{\frac{3}{2}+\frac{2}{3}}}\wedge e^{ \widetilde{\textbf{S}}(\textbf{y})}d\textbf{W}(\textbf{y})
\end{equation}
where $\widetilde{\textbf{S}}$ is a tensorial Gaussian log-correlated noise, inspired by (\ref{eq:K41Field}) and  (\ref{eq:Xscal}), of the form 
\begin{align}\label{eq:Slog}
\widetilde{\textbf{S}}(\textbf{y}) &= \sqrt{\frac{5}{4\pi}}\lambda\int_{|\textbf{y}-\vct{\sigma}|\le L}\left[ \frac{(\textbf{y}-\vct{\sigma})\otimes [(\textbf{y}-\vct{\sigma})\wedge d\textbf{W}(\vct{\sigma})]}{|\textbf{y}-\vct{\sigma}|_\epsilon^{7/2}}\right.\notag \\
& \left.+ \frac{ [(\textbf{y}-\vct{\sigma})\wedge d\textbf{W}(\vct{\sigma})]\otimes(\textbf{y}-\vct{\sigma})}{|\textbf{y}-\vct{\sigma}|_\epsilon^{7/2}}\right]\mbox{ .}
\end{align}
It can be shown that, for instance, the diagonal components correlation is of the form $\langle \widetilde{S}_{11}(\textbf{y}) \widetilde{S}_{11}(\textbf{0}) \rangle \sim \frac{8}{3}\lambda^2 \ln(L/|\textbf{y}|)$ and, for off-diagonal components,  $\langle \widetilde{S}_{12}(\textbf{y}) \widetilde{S}_{12}(\textbf{0}) \rangle \sim 2\lambda^2 \ln(L/|\textbf{y}|)$ in the limit $\epsilon\ll |\textbf{y}| \rightarrow 0$. In the tensorial case, calculations are difficult and the velocity field (\ref{eq:KO62Field}) is expected to be asymptotically multifractal \cite{RobVar08} with a quadratic structure exponent, i.e.  for the longitudinal case  $\partial^2\zeta_q^L/\partial q^2 = -c\lambda^2$. A rigorous  derivation of the constant  $c$ is still missing; we will present {latterly} that numerics show that $\zeta_q^L\approx 1$ and $c\approx 1$ (see the flatness in  Fig. \ref{fig:KO62}(b) and the discussion of Fig. \ref{fig:EstimZetaq}). The intermittency coefficient $\lambda$ is chosen as $\lambda^2 = 0.025$  on empirical grounds \cite{CheCas06,CheCas06_2}. 

\section{Numerical results}
\begin{figure}[t]
\begin{center}
\epsfig{file=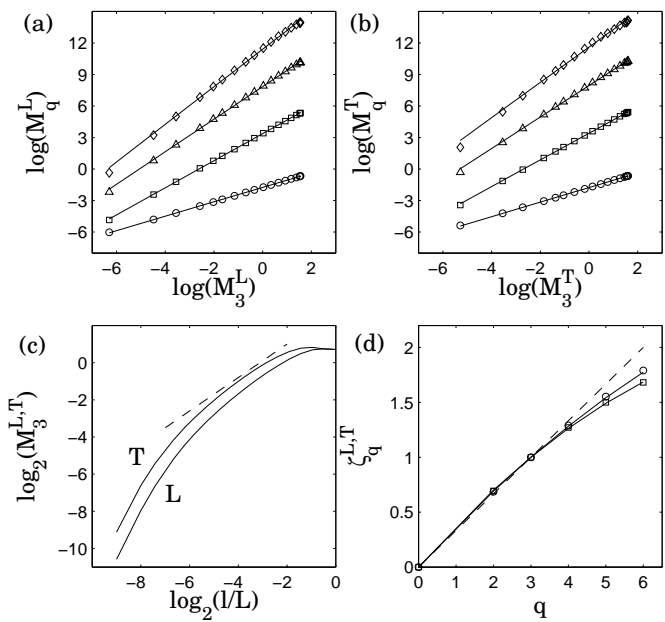,width=8.8cm}
\end{center}
\caption{\label{fig:EstimZetaq} Intermittent characteristics of the process given in (\ref{eq:KO62Field}) for the $N=1024$ case. (a) (resp. (b)) Relative behavior of the longitudinal $M_q^L$ (resp. transverse $M_q^T$) structure functions with respect to the corresponding third order moment $M_3^L$ (resp. $M_3^T$). Orders considered are represented with different symbols: $\circ$ ($q=2$),  $\square$ ($q=4$), $\triangle$ ($q=5$) and $\diamond$ ($q=6$). Straight lines correspond to a least-square fit procedure. Curves are vertically arbitrary shifted for clarity. (c) Scale dependence of the corresponding third order structure function $M_3^L/\sigma^3$ (lower) and $M_3^T/\sigma^3$ (upper), with $\sigma^2=2\langle u_1^2\rangle$. Dashed line corresponds to a unit slope line. (d) Corresponding sets of exponents $\zeta_q^L$ ($\circ$) and $\zeta_q^T$ ($\square$) assuming that the third order structure functions behave as a linear function of the scale in the inertial range, in both the longitudinal and transverse cases. The dashed line corresponds to the K41 prediction $\zeta_q=q/3$ and solid curves to quadratic fits (see text).}
\end{figure}

We display in Fig. \ref{fig:KO62} the results of simulations of the process given by (\ref{eq:KO62Field}) with resolutions $N=256, 512, 1024$. We first see in Fig. \ref{fig:KO62}(a) (resp. (b)) the typical continuous shape deformation of the longitudinal (resp. transverse) velocity increments PDFs characteristic of intermittency and turbulence (see Ref. \cite{CheCas06,CheCas06_2}). The dissymmetry of PDFs in the longitudinal case should be noted. In Fig. \ref{fig:KO62}(c), the obtained RQ-plane is realistic of a fully developed turbulent flow. We reproduce in Fig. \ref{fig:KO62}(d) the skewness and flatness of velocity increments. First Flatness values are much bigger than in the first case and large length scales are populated by intermittency. A rough power law is obtained of slope $-0.1$ showing that $c\approx 1$ in the longitudinal case (see also the following discussion of Fig. \ref{fig:EstimZetaq}). In the same spirit, skewness of longitudinal increments is non-zero at any scale in a realistic way. Finally, let us focus on alignment properties
of vorticity. In Figs. \ref{fig:KO62}(e) and (f) are displayed various PDFs of the cosine of the angle between the vorticity and the eigenframe of the deformation. We see that as the resolution N increases (Fig. \ref{fig:KO62}(e)), the vorticity gets preferentially aligned with the intermediate eigenvector, whereas it gets uncorrelated with the direction of the eigenvector associated
to the most extensive eigenvalue. In Fig. \ref{fig:KO62}(f) is reproduced the alignments properties with the three eigenvectors of the deformation of the $N=1024$ case, which is realistic of a turbulent flow. {We can see here that including the multifractal phenomenology into the linearized form (Eq. (\ref{eq:K41Field})) allows to predict surprisingly both skewness and peculiar alignments of vorticity. A full theoretical explanation of these numerical results is still missing, some analytics in this direction would help to understand these facts.} Let us now focus precisely on the intermittent nature of this velocity field, in particular on a possible difference between longitudinal and transverse velocity increments.
 
On the one hand, the turbulence model proposed in Eq. (\ref{eq:KO62Field}) depends on a single free parameter $\lambda$, the intermittency parameter. On the other hand, two types of intermittency can be identified on a vectorial field, namely the longitudinal and transverse ones. From both experimental and numerical flows, these two types of intermittency seem to be different (see Refs. \cite{CamBen97,GotFuk02} and references therein). As we mentioned, an exact derivation from the model of the corresponding longitudinal $\zeta_q^L$ and transverse $\zeta_q^T$ sets of structure functions exponents is still missing because mainly of the difficulty to deal with the tensorial nature of the matrix exponential. In order to precisely quantify these sets of exponents, we provide a standard fit procedure of the behavior in scale of the structure functions, in a relative fashion in the spirit of Ref. \cite{BenziESS}.

In Figs. \ref{fig:EstimZetaq}(a) and (b), we perform a least-square fit of the relative high orders {longitudinal} and transverse structure functions. This gives access to both the longitudinal $\zeta_q^L$ and transverse $\zeta_q^T$ sets of exponents, assuming for the third order structure a linear behavior with the scale $\ell=|\textbf{r}|$. We indeed see in Fig. \ref{fig:EstimZetaq}(c) that in a first approximation, $M_3^L$ and $M_3^T$ behave as a linear function of the scale. In the sequel, we will thus take $\zeta_3^{L,T}\approx 1$, this could be exactly imposed if the exponent $\frac{3}{2}+\frac{2}{3}$ entering in Eq. (\ref{eq:KO62Field}) is slightly changed. Corresponding structure functions sets of exponents are reproduced in  Fig. \ref{fig:EstimZetaq}(d). One can see that transverse velocity fluctuations are more intermittent than the longitudinal ones, as it has been seen in experiments and numerical simulations \cite{CamBen97,GotFuk02}. The solid curves of Fig. \ref{fig:EstimZetaq}(d) are quadratic functions of $q$:
$\zeta_q^{L,T} = \left( \frac{1}{3}+\frac{3}{2}c^{L,T}\lambda^2\right)q-c^{L,T}\lambda^2\frac{q^2}{2}\mbox{ .}$
Following the fit procedure formerly described, numerics show that $c^L\approx 1$ and $c^T \approx 1.4$, showing that transverse velocity increments are more intermittent than longitudinal ones. These results are consistent with the ones obtained from closures of the dynamics of the velocity gradients \cite{CheMen07b}.

\section{Conclusion and final remarks}
As a conclusion, based on prior works \cite{CheA2,RobVar08}, we have built an incompressible skewed intermittent velocity field that reproduces the main characteristics of 3D fully-developed turbulence in the inertial range. This includes several non trivial geometrical properties (RQ-plane and the preferential alignments of vorticity), a non-vanishing skewness of the longitudinal velocity increments and a realistic intermittent picture of longitudinal and transverse velocity fluctuations.  The model provided in Eqs. (\ref{eq:KO62Field}) and (\ref{eq:Slog}) can thus be seen as a stochastic representation of the fields described by Kolmogorov and Obouhkov \cite{Kolmo62,Obu62}.  To do so, we included the multifractal phenomenology to the Euler mechanics at short-time. Several remarks can be made at this stage. First, this theory contains a free parameter $\lambda$ chosen to be consistent with experimental findings. It would be very interesting to find constraints able to lead to a direct determination of $\lambda$. {In this work we make use of some well known facts, like RFD closure and intermittency, to build a rather realistic stationary model of turbulent velocity field. But we make no use of the very dynamics of the system. We may imagine that such a field is close in some sense to an invariant measure for Euler or Navier -Stokes dynamics and that this would help us to fix the value of the unknown parameter $\lambda$. Hence, time correlations and energy (or enstrophy) budgets are not, at this stage, predicted.} {Finally}, this model could also {help} in the understanding of the physics of the pressure Hessian  \cite{CheA2,Luthi09}. We leave these aspects for future investigations.

\acknowledgments
We thank B. Castaing, Y. Gagne and K. Gawedzki for fruitful discussions, S.G. Roux for numerical advices and the PSMN (H. Gilquin) for computer time.

\end{document}